\documentclass[11pt,fullpage, a4paperr]{article}
\usepackage{amssymb,amsmath}
\usepackage[left=1in,top=1in,right=1in,bottom=1in,nohead]{geometry}
\usepackage{graphicx}
\usepackage{epsfig}
\usepackage{caption}
\usepackage{float}
\usepackage{setspace}
\usepackage{natbib}
\usepackage{hyperref}
\usepackage{lscape}
\usepackage{longtable}
\usepackage{tabularx}
\usepackage {tikz}
\usepackage{ctable}
\usepackage{booktabs}
\usepackage{lscape}
\usepackage{float}
\usepackage{bibentry}
\usepackage{bbm}
\usepackage{siunitx,booktabs}
\usepackage{threeparttable}
\usepackage{multirow}
\usepackage{comment}
\usepackage{threeparttable}

\usepackage{amsthm}
\usetikzlibrary{calc}

\usetikzlibrary{arrows.meta}
\hypersetup{
	colorlinks=true,
	citecolor=blue}

\author{Natsuki Arai\thanks{Department of International Business, National Chengchi University, Taipei, Taiwan
} \footnote{Corresponding author. Address: Department of International Business, National Chengchi University, No. 64, Sec. 2, Tz-nan Rd., Wenshan, Taipei 116, Taiwan. E-mail: \textsc{\href{narai@nccu.edu.tw}{narai@nccu.edu.tw}}.}
 \and Shian Chang\footnotemark[1] 
 \and Biing-Shen Kuo\footnotemark[1]}
\title{Introductory Economics: Gender, Majors, and Future Performance}
\date{\today}

\begin{document}
\setcounter{page}{1}
\maketitle
 \thispagestyle{empty}
\begin{abstract}
By investigating the exam scores of introductory economics in a business school in Taiwan between 2008 and 2019, we find three sets of results: First, we find no significant difference between genders in the exam scores. Second, students' majors are significantly associated with their exam scores, which likely reflects their academic ability measured at college admission. Third, the exam scores are strong predictors of students' future academic performance.
\end{abstract}

\bigskip
\textbf{Keywords}: Undergraduate Education, Introductory Economics, Gender, Major, Future Performance

\bigskip
\textbf{J.E.L. codes}: A22, I23, J16


\newpage
\setcounter{page}{1}

\section{Introduction}
The lack of diversity in the economics profession across  different  genders  has  been  widely recognized in the literature. Recently, it has become even more problematic because progress  in  the economics profession has been slower than in other disciplines, such as science,  technology,  engineering,  and math (STEM) fields, which are typically perceived as male domains. \cite{LS2019} claimed that ``women's progress in academic economics has slowed, with virtually no improvement'' and \cite{WSN2019} showed that the proportion of female economists with Ph.D. in academia is 23\%, fewer than 30\% in the US government. Various  studies have  been conducted to explain the underrepresentation of women in economics, and two different views are presented: Earlier  studies, such as \cite{DR1997}, suggested  that  math  intensity is crucial,  while more recent studies, including \cite{CGKW2014}, indicated that numerous barriers during researchers' careers are vital. As the policy implication is substantially different depending on these views, investigating which view has more robust empirical support is crucial.
 
 This paper analyzes whether performance in introductory economics is associated with the individual characteristics of students by using the institutional data from the business school of National Chengchi University (NCCU) in Taiwan. There are several advantages to using this dataset. First, the data covers the panel of students between 2008 and 2019, more than ten years, which offers a sufficiently large sample for the analysis. Second, the comparability of exam scores and students across different years is ensured due to the following institutional features: (1) introductory economics is required for all majors in the business school, which likely mitigates the issue of selection bias, (2) introductory economics is based on a standardized syllabus, which was unchanged during the sample period, and (3) NCCU is one of the most prestigious institutions in Taiwan for social sciences.
 
This paper employs standard regression analysis to determine three sets of results. First, the association between gender and performance in introductory economics is not statistically significant. Second, the association between exam scores and students' majors is statistically significant in most cases. The signs and magnitudes of the coefficients are consistent with the rankings of majors at college admission, suggesting that performance in introductory economics reflects students' academic ability measured at the admission. Third, the exam scores in introductory economics are a strong predictor of students' academic performance in the future.
 
The main contribution of this paper is to find empirical evidence that students' academic ability at admission is more critical than gender to explain performance in introductory economics. Given that economics is a highly integrated discipline, requiring math skills and knowledge in various social science fields, it is not surprising that the results underscore the importance of students' academic ability. These results also suggest that investigating the choice of study at graduate schools, outcome of the job market, and promotion of assistant professors, is likely to be more important for the underrepresentation of women in economics. This implication is consistent with recent literature's claims that various cultural and institutional barriers are crucial, including gender stereotyping among economists by \cite{Wu2018}, implicit attitudes and institutional practices by \cite{BR2016}, and lack of role models by \cite{RB2002} and \cite{PS2020}.


\section{Background}\label{sec:Background}
We analyze the exam scores of undergraduate introductory economics at the business school of NCCU and the academic awards given by NCCU between 2008 and 2019. This section describes the structure of introductory economics, coverage and format of the exams, and academic awards.\footnote{Unfortunately, data regarding detailed backgrounds of students or their careers after graduation is not available.}

\subsection{Structure of the Course}
Introductory economics is an annual course covering microeconomics in the first semester and macroeconomics in the second. The course is primarily designed for first-year students, and students with all eight majors at the business school are required to take: (1) International Business, (2) Money and Banking, (3) Accounting, (4) Statistics, (5) Business Administration, (6) Management Information Systems, (7) Finance, and (8) Risk Management and Insurance. The students are split into several sections, each of which is taught by different instructors. The instructors can design their sections, including whether they teach in Mandarin or English. However, students take an identical final exam in English at the end of each semester. The assignment of sections is primarily based on majors due to the scheduling issues. 

Table \ref{tab:departments} summarizes the number of students with different gender and majors for the whole sample period. The total observation is 15,731, and the proportion of female students is larger than male students for most majors, except for statistics and the school of science. Approximately 30\% of students are from non-business schools, including the schools of Arts and Foreign Languages.

\subsection{Coverage and Format of the Exam}
\begin{figure}
	\centering
	\includegraphics[width=0.9\textwidth]{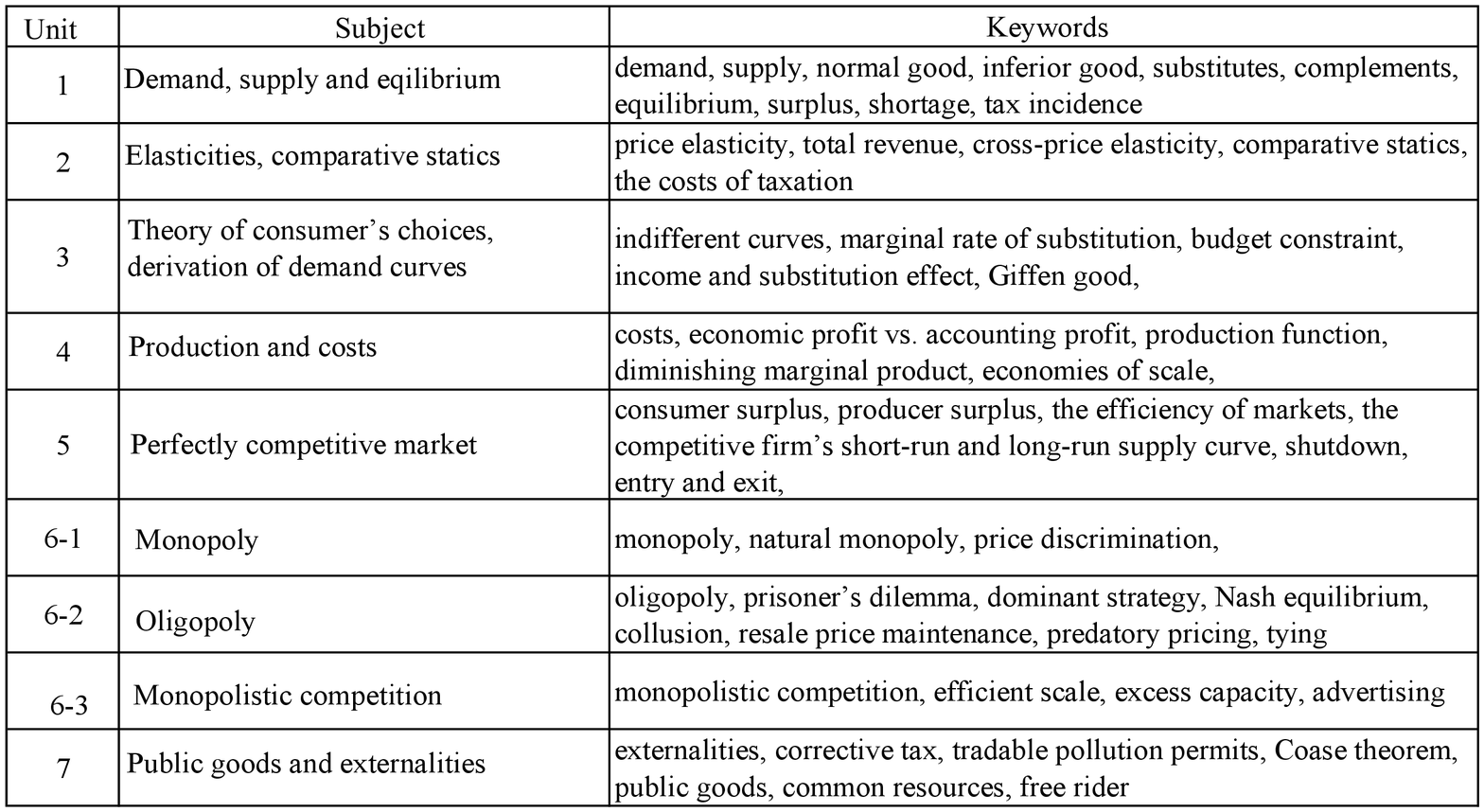}
	\caption{Coverage in Microeconomics (First Semester)}
	\label{fig:micro}
\end{figure}

\begin{figure}
	\centering
	\includegraphics[width=0.9\textwidth]{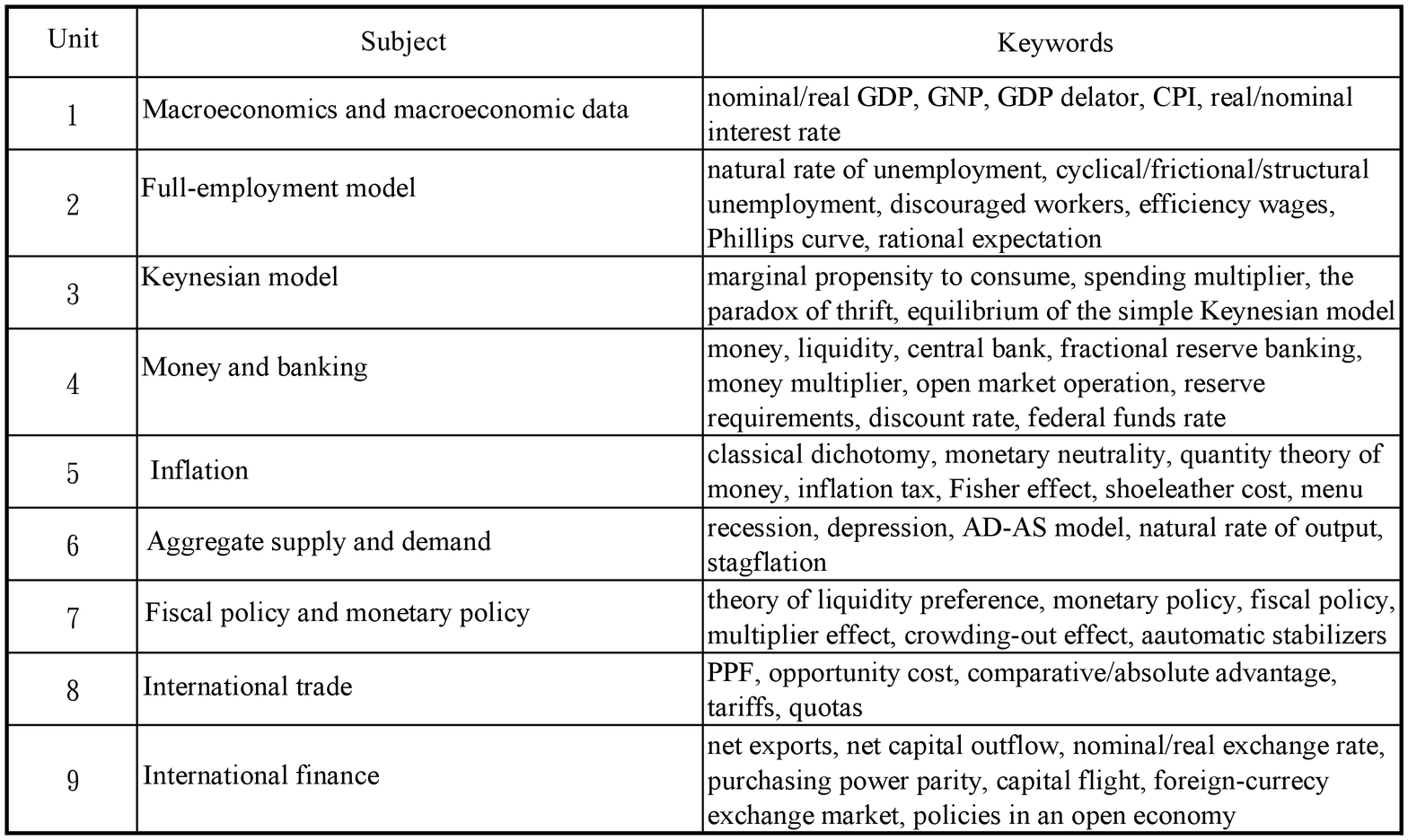}
	\caption{Coverage in Macroeconomics (Second Semester)}
	\label{fig:macro}
\end{figure}

Various topics in microeconomics and macroeconomics are covered in the course, the details of which are listed in Figures \ref{fig:micro} and \ref{fig:macro}, respectively. The materials follow standard economics textbooks, such as \cite{Mankiw2018} and \cite{HO2018}. 
The final exam consists of  50 multiple-choice questions in English. 

Table \ref{tab:mean} provides the summary statistics of the exam. Although the average score across all semesters is 61.0 with a standard deviation of 17.0, the average score of microeconomics is 62.2 and slightly higher than the average score of macroeconomics, 59.6. The average number of students is 684 per semester.

\subsection{Academic Award}
\begin{figure}
	\centering
	\includegraphics[width=13cm]{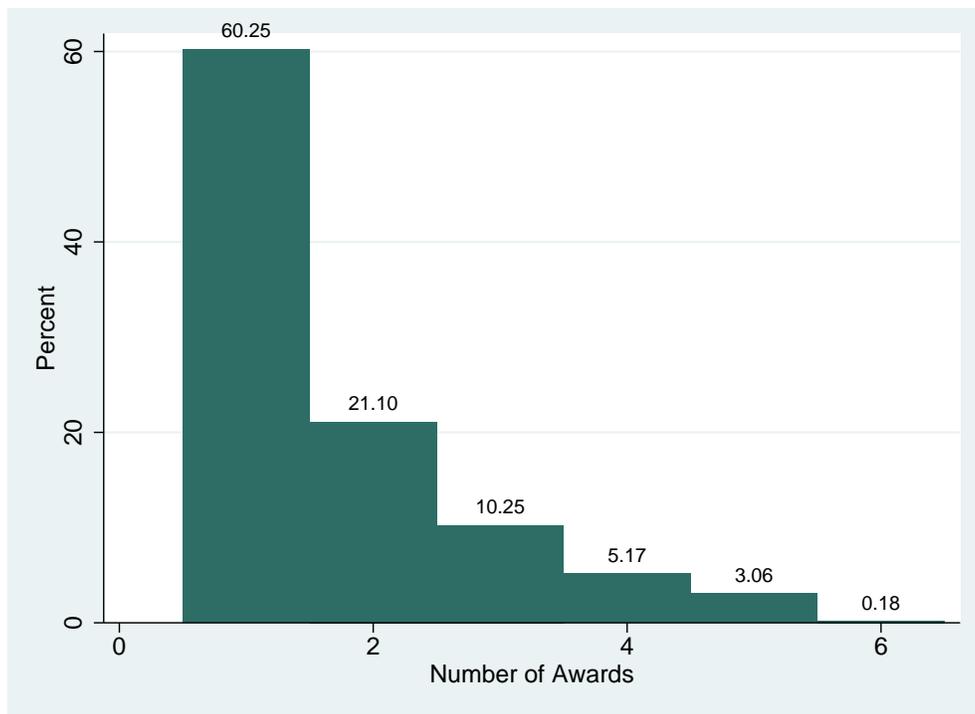}
	\caption{Distribution of the Number of Academic Awards (Conditional on Receiving an Award)}
	\label{fig:awardHist}
\end{figure}

Every semester, the three students who earn the highest average scores in each major and year receive an official academic award from NCCU. In other words, tweleve students in each major per semester receive the award. We use this academic award as a measure of future academic performance. Table \ref{tab:mean} also provides the summary statistics showing that the award is reasonably competitive---only 13\% of students have received it. Figure \ref{fig:awardHist} shows the conditional distribution of the number of awards, and more than half of them receive the award only once. 

\section{Econometric Model}\label{sec:Model}
\subsection{Individual Characteristics}

We employ a regression model to analyze the association between exam scores and individual characteristics as follows:
\begin{equation}
      Z^{Score}_{i,t}  =  \alpha + \beta_{1}D^{Female}_{i}+\sum_{m=1}^M\beta_{2m}D^{Major}_{i}+\gamma X_{i,t}+\varepsilon_{i,t}. \label{eq:testScore}
\end{equation}
The dependent variable, $Z^{Score}_{i,t}$ is the standardized exam score of student $i$ at year $t$. $D^{Female}$ and $D^{Major}$ are the dummy variables for the female students and different majors, respectively, where $M$ denotes the total number of majors. Since the number of students from other schools is much fewer than those from the business school, they are grouped with their schools instead of majors. Control variables $X_{i,t}$ include the dummy variables for the second semesters (macroeconomics), the years of the exams, instructors, and the years of the students. The model is estimated by pooled OLS using the whole sample and by the random effect model using the sample of students who take both micro and macro exams.

\subsection{Future Academic Performance}
We also analyze the association between exam scores in introductory economics and students' future academic performance at NCCU. We construct the dummy variable based on academic awards as follows:

\begin{equation}
	D^{Award}_{i}\equiv
	\begin{cases}
		1 & \text{if} \  \mathbbm{1}_{Award, \tau}=1 \text{ for } \tau>t  \  \text{ such that } Score_{i,t} \neq \emptyset,\\
		0 & \text{otherwise},\\
	\end{cases}
	\label{eq:awardDef}
\end{equation}
where $Score_{i,t}$ denotes the exam score of student $i$ at year $t$.\footnote{For the students who take both micro and macro exams, we use the higher score.} In other words, this dummy variable indicates whether the student receives the academic award after they take introductory economics class. Then, we run the following linear regression:

\begin{equation}
	D^{Award}_{i}=\alpha+\beta_1 Z^{Score}_{i,t} +\beta_{2}D^{Female}_{i}+\beta_{3}( Score_{i,t} \times D^{Female}_{i}) +\gamma X_{i,t}+\varepsilon_{i,t}. \label{eq:Award}
\end{equation}
 The control variable $X_{i,t}$ include the dummy variables for the years of the exam, instructors, students' majors and years. The interaction term of gender and exam score is included to capture the difference in comovement between exam score and award across genders. 

\section{Results}\label{sec:Res}
\subsection{Individual Characteristics}
Table \ref{tab:result} shows the results based on pooled OLS and random effect model, for the full sample and the subsample of first-yer students. Since the dummy variables are normalized with male first-year students majoring in international business, the estimated coefficients indicate the relative performance against this benchmark group. The estimates for the dummy variables of female students are statistically insignificant in all cases, though they are estimated to be positive. On the other hand, the estimates for the dummy variables of majors are statistically significant except for the accounting majors. While students majoring in money and banking and finance perform significantly better, with magnitudes between 0.07 and 0.14 for different specifications, students majoring in statistics, business administration, management information systems, and risk management and insurance tend to perform significantly worse, with magnitudes between -0.34 and -0.16. The estimates for other schools are significantly negative, except for the law school. 

\begin{figure}
	\centering
	\includegraphics[width=13cm]{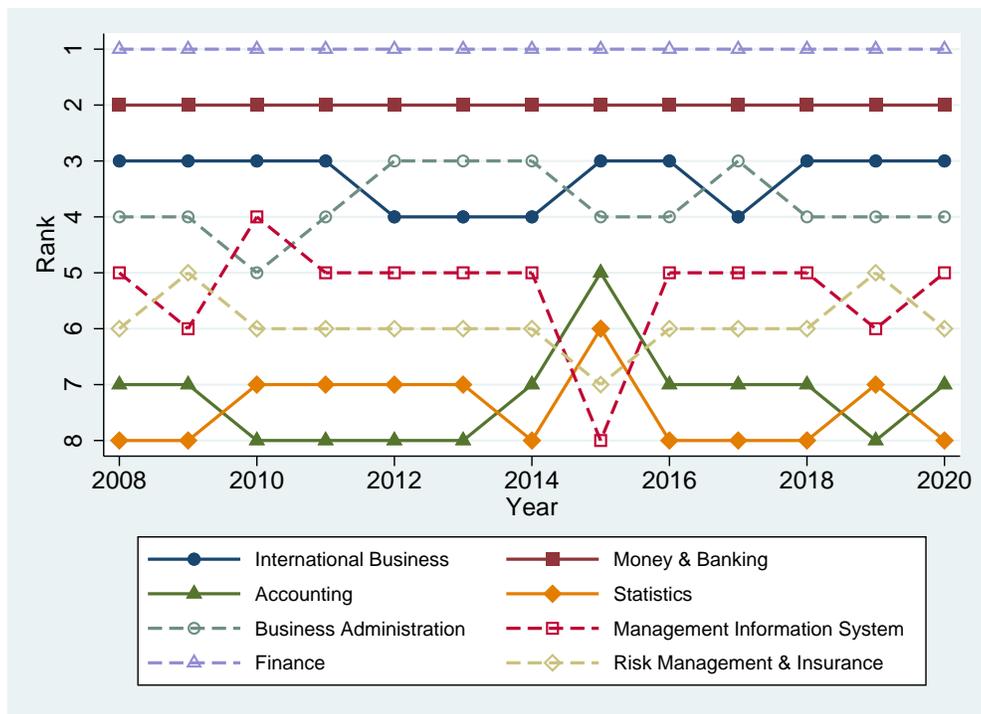}
	\caption{Rankings of Majors Based on the Exam Scores for Admission}
	\label{fig:RankDep}
\end{figure}

These results likely reflect the academic ability measured at admission to NCCU. Figure \ref{fig:RankDep} shows the time series of the rankings of majors sorted with the required scores of the national exam for college admission in Taiwan.\footnote{The national exam plays an important role in college admission in Taiwan. For a comprehensive analysis of college admissions in Taiwan, see \cite{LKLL2020} and \cite{LLLWK2020}.} As indicated in the figure, admission to the finance and money and banking majors requires consistently higher scores than those for international business majors. On the other hand, the rankings of other majors are lower than international business in most cases, consistent with the significantly negative estimates in the regression. This strong association between the estimates the rankings of majors at admission suggests that academic ability measured at admission has strong predictive power for the performance in introductory economics. These results are similar to those of \cite{BJ2004}, claiming that basic math skills are critical for the performance in introductory economics.

\subsection{Future Academic Performance}
Table \ref{tab:result2} presents the results based on different specifications and estimations. The results show that the association between the exam scores of introductory economics and future academic performance is positive and statistically significant. One standard deviation in the exam score of introductory economics is associated with a 10 percentage points increase in the probability of receiving the academic award in the linear probability model---it substantially increases the probability of receiving the award by 76\% relative to the mean. The results are similar even when the model is estimated by nonlinear specifications, such as the logit and probit models, and when we use the number of awards received as the regressand, finding significantly positive associations between exam scores and future academic performance. The estimated coefficients for the interaction term of female and exam score are positive and statistically significant in the linear probability model, which may reflect the higher mean and lower variance of female students' grades as discussed in \cite{OLJN2018} and \cite{VV2014}. However, the estimates are not statistically significant for logit and probit models.

These results imply that students' academic ability measured by the introductory economics exam have strong predictive power for their future academic performance. 

\section{Conclusion}\label{sec:Con}
This paper analyzes the relationship between performance in introductory economics and individual characteristics and finds that there is no substantial difference between genders. The discrepancies between majors are, however, statistically significant, reflecting the academic ability measured at college admission. The additional analysis finds a significant association between exam scores in introductory economics and future academic performance, suggesting that the introductory economics exams are a reliable measure of students' academic ability.

These results have an important policy implication for debates on the underrepresentation of women in the economics profession. Consistent with findings in recent literature, this paper finds little evidence of a difference in academic ability between genders in economics. Therefore, various interventions to remove barriers at different stages of researchers' careers, as in \cite{Buckles2019}, may be a more successful approach to promoting gender diversity in economics profession.

\newpage
\section*{Acknowledgments}
The authors are grateful to Nobuhiko Nakazawa and Kyungmin Kang for their helpful comments and to Alice Lee for her excellent administrative support. Arai gratefully acknowledges financial support from the Taiwanese Ministry of Science and Technology's grant 108-2410-H-004-009. All errors are the sole responsibility of the authors.
\clearpage

\newpage
\bibliographystyle{aer}
\bibliography{ProposalRef}
\clearpage

\clearpage
\newpage

\clearpage
\newpage
\clearpage
\newpage

\clearpage
\newpage

\begin{table}[t]
	\centering
	\begin{tabular}{m{1mm}m{3mm}l r@{\hspace{2mm}}l r@{\hspace{2mm}}l r} \toprule
		\multicolumn{3}{l}{Majors}& 	\multicolumn{2}{c}{Female} & \multicolumn{2}{c}{Male}& Total\\ \midrule
		\multicolumn{3}{l}{Business School}  &&&&&\\
		&&International Business  &  1,137& (64\%) &  636 &(36\%) &   1,773 \\
		&&Money and Banking       &  637  & (56\%) &  506 &(44\%) &   1,143 \\
		&&Accounting   			  &  1,284& (64\%) &  737 &(36\%) &   2,021 \\
		&&Statistics              &   477 & (44\%) &  606 &(56\%) &   1,083 \\
		&&Business Administration &  1,290& (60\%) &  872 &(40\%) &   2,162 \\
	    &&Management Information Systems          &  238& (49\%) & 246& (51\%)   &     484 \\
		&&Finance                 & 575 &(58\%)    &  420 &(42\%) &   995 \\
		&&Risk Management and Insurance           &  581 &(55\%) &   482 &(45\%) &   1,063 \\
		&\multicolumn{2}{l}{Business School Total}& 6,219 &(58\%)& 4,505& (42\%) & 10,724 \\ \midrule
		\multicolumn{3}{l}{Other Schools}  & &&&&\\
		&& Arts                   &  378 &(55\%)   & 305& (45\%)  &    683 \\
		&& Social Science         &  773 &(58\%)   & 551 &(42\%)  &   1,324 \\
		&& Communication          & 377& (71\%)    & 154& (29\%)  &     531 \\
		&& Foreign Language and Literature & 1,138& (74\%)&  403 &(26\%)   &   1,541 \\
		&& Law &     177& (54\%) &     149& (46\%)    &     326 \\
		&& Science   &   185& (31\%)&  417& (69\%)   &     602 \\
		&\multicolumn{2}{l}{Other Schools Total} & 3,028 & (60\%) & 1,979 & (40\%)& 5,007 \\ \midrule
			\multicolumn{3}{l}{Total} &  9,247& (59\%)&    6,484 &(41\%)&    15,731\\  \bottomrule
	\end{tabular}
	\caption{Number of  Samples from Different Majors and Gender}
	\label{tab:departments}
\end{table}
\clearpage
\newpage
\begin{table}[t]
	\centering
	\begin{threeparttable}
	\begin{tabular}{l l *{3}{S[table-format=3.2] }} \toprule
		& & {Female} & {Male} & {Total}\\
		\midrule
		Exam Score:& Overall Average & 61.62 & 60.11 & 60.99 \\
		&& {(}16.49{)} & {(}17.57{)} & {(}16.96{)}\\
		& Average in Microeconomics& 62.74 & 61.44 & 62.20\\
		& &  {(}15.73{)} &  {(}17.05{)} &  {(}16.30{)}\\
		&Average in Macroeconomics & 60.30 & 58.55 & 59.57\\
		&  & {(}17.26{)} &  {(}18.03{)} &  {(}17.60{)}\\ 
		Academic Award:& Average Number & 0.25 & 0.17 & 0.22 \\ 		
		&& {(}0.73{)} & {(}0.61{)} & {(}0.68{)}\\
		 & Probability to Receive & 0.15 & 0.10 & 0.13 \\
		& & {(}0.35{)} & {(}0.30{)} & {(}0.33{)} \\

		\multicolumn{2}{l}{Number of Observations per Semester} & 402.04 & 281.91 & 683.96\\ \bottomrule
	\end{tabular}
	\caption{Summary Statistics}
	\label{tab:mean}
	   \begin{tablenotes}
		\footnotesize 
		\item [a.] This table shows the summary statistics of the series for different gender. Numbers in parenthesis show the standard deviation of the correpsonding series.
	\end{tablenotes}
\end{threeparttable}
\end{table}
\clearpage
\newpage

\begin{table}[t]
\begin{center}
	 \begin{threeparttable}
\small 
\begin{tabular}{m{3mm}  l*{4}{S[table-format=3.2] }}
	\toprule
\multicolumn{2}{l}{Sample}& \multicolumn{2}{c}{Full Sample} & \multicolumn{2}{c}{First-Year Students}\\
&& {(1)} & {(2)}&  {(3)} & {(4)}\\
\multicolumn{2}{l}{Estimation}& {Pooled OLS} & {Random Effect} & {Pooled OLS} & {Random Effect} \\
 \midrule
\multicolumn{2}{l}{Regressors}& &&& \\ 
&Constant & 0.47$^{***}$ & 0.41$^{***}$& 0.49$^{***}$ & 0.42$^{***}$\\ 
&              &{(}0.04{)} & {(}0.05{)} & {(}0.04{)}&{(}0.05{)}\\			
&Female & 0.02 & 0.01  & 0.02 & 0.01\\ 
&              &{(}0.01{)} & {(}0.02{)} & {(}0.02{)}&{(}0.02{)}\\			
\multicolumn{2}{l}{Majors} &&&&\\
&Money and Banking& 0.13$^{***}$& 0.11$^{**}$ & 0.14$^{***}$ & 0.12$^{**}$\\ 
&              &{(}0.04{)} & {(}0.05{)} & {(}0.04{)}&{(}0.05{)}\\			
&Accounting &  0.01 & 0.02 & 0.00 & 0.01\\ 
&              &{(}0.03{)} & {(}0.04{)} & {(}0.03{)}&{(}0.04{)}\\		
&Statistics & -0.21$^{***}$& -0.21$^{***}$ & -0.20$^{***}$ & -0.20$^{***}$\\ 
&              &{(}0.03{)} & {(}0.04{)} & {(}0.03{)}&{(}0.04{)}\\		
&Business Administrations& -0.28$^{***}$ & -0.27$^{***}$ & -0.28$^{***}$ & -0.27$^{***}$\\ 
&              &{(}0.03{)} & {(}0.04{)} & {(}0.03{)}&{(}0.04{)}\\		
&Management Information Systems &  -0.34$^{***}$ & -0.31$^{***}$ & -0.24$^{***}$ & -0.25$^{***}$\\ 
&              &{(}0.04{)} & {(}0.06{)} & {(}0.05{)}&{(}0.06{)}\\		
&Finance& 0.08$^{**}$ & 0.10$^{**}$ & 0.07$^{**}$ & 0.09$^{*}$\\ 
&              &{(}0.03{)} & {(}0.05{)} & {(}0.03{)}&{(}0.05{)}\\		
&Risk Management and Insurance& -0.16$^{***}$ & -0.16$^{***}$ & -0.16$^{***}$ & -0.16$^{***}$\\ 
&              &{(}0.03{)} & {(}0.04{)} & {(}0.03{)}&{(}0.04{)}\\		

\multicolumn{2}{l}{Other Schools} &&&&\\
&Arts&-0.27$^{***}$ & -0.24$^{***}$ & &\\ 
&              &{(}0.04{)} & {(}0.05{)} &&\\		
&Social Science &-0.21$^{***}$& -0.14$^{***}$ &  &\\ 
&              &{(}0.03{)} & {(}0.05{)} &&\\		
&Communication &-0.14$^{***}$ & -0.12$^{**}$ & &\\ 
&              &{(}0.04{)} & {(}0.06{)} &&\\		
&Foreign Language &-0.18$^{***}$ & -0.15$^{***}$ & &\\ 
&              &{(}0.03{)} & {(}0.05{)} &&\\		
&Law &0.01 & 0.09 & &\\ 
&              &{(}0.05{)} & {(}0.07{)} &&\\		
&Science & -0.37$^{***}$ & -0.36$^{***}$ & &\\ 
&              &{(}0.04{)} & {(}0.06{)} &&\\		
&Average& & & -0.20$^{***}$& -0.14$^{***}$\\ 
&    &&          &{(}0.03{)} & {(}0.04{)}\\		
\multicolumn{2}{l}{Fixed Effect} &&&&\\
&Semester&{Yes}&{Yes}&{Yes}&{Yes}\\
&Year&{Yes}&{Yes}&{Yes}&{Yes}\\
&Instructor&{Yes}&{Yes}&{Yes}&{Yes}\\
&Student's Year&{Yes}&{Yes}&&\\

& & & &\\
\multicolumn{2}{l}{Wald Statistic}& 8305.30$^{***}$ &  4045.83$^{***}$& 5583.46$^{***}$ & 3002.88$^{***}$ \\ 
\multicolumn{2}{l}{Adjusted $R^{2}$ / Between $R^{2}$ }& 0.32 & 0.36 & 0.29 & 0.33\\ 
\multicolumn{2}{l}{Number of Observations} & 15731 & 13546  & 12680 & 11408 \\ \bottomrule
\end{tabular}
	\caption{Regression Analysis of Exam Scores} \label{tab:result}
   \begin{tablenotes}
	\footnotesize 
	\item [a.] This table shows the estimates analyzing the standardized exam scores described in Equation \eqref{eq:testScore}. Dummy variables are normalized with male first-year students majoring in international business.
	\item [b.] Heteroscedascitity-robust standard errors are reported in parentheses and $^{*}$, $^{**}$, and $^{***}$ indicate the significance at the 10\%, 5\%, and 1\% level, respectively. 
\end{tablenotes}
\end{threeparttable}
\end{center}
\end{table}
\clearpage
\newpage
\begin{table}[t]
	\begin{center}
		\begin{threeparttable}
			\small 
			\begin{tabular}{m{5mm} l*{4}{S[table-format=3.2] }}\toprule
			 	Regressand&& \multicolumn{3}{c}{Prob. of Award} & {Num. of Awards}\\
				&& {(1)} & {(2)}&  {(3)} & {(4)}  \\ 
			  \multicolumn{2}{l}{Estimation} &{Linear Prob.} & {Logit} &  {Probit} & {Linear Prob.}\\	\midrule
			\multicolumn{2}{l}{Regressors}& &&& \\ 
				&Constant & 0.06$^{***}$ & 0.08$^{***}$  & 0.09$^{***}$ & 0.09$^{**}$\\ 
	        	&           &{(}0.02{)} & {(}0.00{)} & {(}0.00{)}&{(}0.04{)}\\				
				&Exam Score& 0.10$^{***}$ & 0.10$^{***}$ & 0.11$^{***}$ & 0.19$^{***}$\\ 
					  	    & & {(}0.01{)} & {(}0.01{)} & {(}0.01{)}&{(}0.01{)}\\
				&Female&  0.04$^{***}$      & 0.04$^{***}$ & 0.04$^{***}$ & 0.06$^{***}$\\ 
				& 			& {(}0.01{)} & {(}0.01{)} & {(}0.01{)}&{(}0.01{)}\\
				& Female $\times$ Exam Score & 0.03$^{***}$ & ${-}$0.00$^{}$  & 0.00$^{}$ & 0.06$^{***}$\\ 
				& 				 & {(}0.01{)} & {(}0.01{)} & {(}0.01{)}&{(}0.02{)}\\
				\multicolumn{2}{l}{Fixed Effect} &&&&\\
				&Year&{Yes}&{Yes}&{Yes}&{Yes}\\
				&Instructor&{Yes}&{Yes}&{Yes}&{Yes}\\
				&Major&{Yes}&{Yes}&{Yes}&{Yes}\\
				&First-Year Students&{Yes}&{Yes}&{Yes}&{Yes}\\
				& & & &\\
				\multicolumn{2}{l}{Mean of Regressand}& 0.13 & 0.13 & 0.13 & 0.22 \\
				\multicolumn{2}{l}{Wald Statistic}& 864.52$^{***}$ & 723.99$^{***}$ & 710.85$^{***}$ & 576.09$^{***}$ \\ 
				\multicolumn{2}{l}{Adjusted $R^2$ / Pseudo $R^{2}$ }& 0.10 & 0.15  & 0.15 & 0.09\\
				\multicolumn{2}{l}{Number of Observations} & 8943 & 8943 & 8943 & 8943 \\  \bottomrule
			\end{tabular}
			\caption{Regression Analysis of Academic Awards} \label{tab:result2}
			\begin{tablenotes}
				\footnotesize 
				\item [a.] This table shows the results analyzing the probability or number of academic awards described in Equation \eqref{eq:Award} based on different estimations. Dummy variables are normalized with male first-year students majoring in international business.
				\item [b.] The table lists the estimates for the linear probability model and corresponding marginal treatment effects for the logit and probit models.
				\item [c.] Heteroscedascitity-robust standard errors are reported in parentheses and $^{**}$ and $^{***}$ indicate the significance at the 5\% and 1\% level, respectively. 
			\end{tablenotes}
		\end{threeparttable}
	\end{center}
\end{table}

\end{document}